\documentclass[letterpaper,10pt]{article} 
\usepackage{opticameet3} %% use version 3 for proper copyright statement

\newcommand\authormark[1]{\textsuperscript{#1}}

\usepackage{amsmath,amssymb}
\usepackage[colorlinks=true,bookmarks=false,citecolor=blue,urlcolor=blue]{hyperref} %pdflatex
\usepackage{pgfplots}
\pgfplotsset{compat=1.18}
\usepackage{tikz}

\usepgfplotslibrary{fillbetween}
\usepgfplotslibrary{colormaps}
\usepgfplotslibrary{colorbrewer}
\usepgfplotslibrary{groupplots}
\usepgfplotslibrary{external} 
\usetikzlibrary{spy}
\usetikzlibrary{arrows.meta}
\usetikzlibrary{arrows}
\usetikzlibrary{fpu}
\usetikzlibrary{shapes.geometric}
\usetikzlibrary{external}

\usepackage{xcolor}
\usepackage{multicol}

\definecolor{grad-1}{HTML}{C7B3CC}
\definecolor{grad-2}{HTML}{A7ABC7}
\definecolor{grad-3}{HTML}{87A3C2}
\definecolor{grad-4}{HTML}{669ABC}
\definecolor{grad-5}{HTML}{4692B7}
\definecolor{grad-6}{HTML}{268AB2}

\definecolor{green1}{RGB}{27,158,119}
\definecolor{orange1}{RGB}{217,95,2}
\definecolor{purple1}{RGB}{117,112,179}
\definecolor{pink1}{RGB}{231,41,138}
\definecolor{lightgreen1}{RGB}{102,166,30}
\definecolor{gray1}{RGB}{102,102,102}
\definecolor{gray2}{RGB}{130,130,130}

\definecolor{o-band}{HTML}{a8d6f4}
\definecolor{e-band}{HTML}{ffbbd5}
\definecolor{s-band}{HTML}{c6afe9}
\definecolor{c-band}{HTML}{8bdebf}
\definecolor{l-band}{HTML}{ffcaa6}

\begin{document}

\title{%
Ultra-Wideband Transmission Systems From an Energy Perspective: Which Band is Next?
% Analysis of Ultra-Wideband Transmission Systems From an Energy Perspective

% Watt band\\next?
% Ultra-Wideband Transmission Systems From an Energy Perspective
}

% \author{Author name(s)}
% \address{Author affiliation and full address}
% \email{e-mail address}
%%Uncomment the following line to override copyright year from the default current year.
%\copyrightyear{2024}

\author{Ronit~Sohanpal\authormark{*,1}, Mindaugus~Jarmolovi\v{c}ius\authormark{1}, Jiaqian~Yang\authormark{1}, Eric~Sillekens\authormark{1}, Romulo~Aparecido\authormark{1}, Vitaly~Mikhailov\authormark{2},
Jiawei~Luo\authormark{2}, 
David~J.~DiGiovanni\authormark{2},
Ruben~S.~Luis\authormark{3}, 
Hideaki~Furukawa\authormark{3}, 
Robert~I.~Killey\authormark{1} and Polina~Bayvel\authormark{1}}

\address{\authormark{1}Optical Networks Group, UCL (University College London), London, UK \\
\authormark{2}Lightera Labs, Somerset, NJ 08873, USA\\
\authormark{3}National Institute of Information and Communications Technology, Koganei 184-8795, Japan}

\email{\authormark{*}ronit.sohanpal@ucl.ac.uk} %% email address is required

\begin{abstract} %35 words
% We measure the power efficiency of state-of-the-art OESCL-band amplifiers. Using the integral GN model, we show that OESCL-band systems can achieve 2.98$\times$ additional throughput for +48\% higher energy-per-bit compared to CL-band transmission only at 1000km.
Measuring the power efficiency of the state-of-the-art OESCL-band amplifiers, we show that 1000~km OESCL-band systems can achieve 2.98$\times$ greater throughput for +48\% higher energy-per-bit compared to CL-band transmission only.
\end{abstract}

% We experimentally investigate the nonlinearity compensation performance of frequency combs and independent lasers sources at 15 and 49.5~GBd over 7796 km, observing 0.5~dB gain using combs over unsynchronised laser sources.

\vspace{-0.01cm}
\section{Introduction}
\vspace{-0.1cm}

Ultra-wideband (UWB) communications have been widely investigated as a proposed solution to meet growing capacity demands by using the full transmission window of standard single-mode silica fibres. A variety of transmission bands have been investigated as potential candidates to expand beyond the conventional and long-wavelength C- and L-bands, up to the entire OESCLU-band transmission window \cite{Hamaoka2025,Puttnam2025}. The biggest enabler of UWB transmission is the development of novel optical amplifiers. Although erbium-doped fibre amplifiers (EDFAs) are well-established for the C- and L-bands, doped fibre technologies for other bands are significantly less mature. However, recent developments in bismuth DFAs (BDFAs) have greatly enhanced the capability for O- and E-band transmission \cite{Mikhailov2022,Donodin2025}. Most UWB transmission research has focussed primarily on maximising throughput and reach, both of which require amplifiers providing sufficient gain, gain flatness and low noise figure within each band. To date, %there has been little research into 
the net power efficiency of these novel amplifiers, i.e., the electrical power needed to attain the required optical characteristics, has not been investigated. For UWB systems to progress towards deployment, it is not sufficient to consider only the band-wise throughput -- the energy-per-bit of each newly added band must be taken into account. Reducing link power consumption is desirable for the potential environmental, power delivery and cost-per-bit benefits. Previous studies in this area have compared lumped and distributed Raman amplification (DRA) for C-band and SCL-band systems, showing that DRA, widely considered to be more power-hungry than lumped amplifiers, can in fact improve system energy-per-bit in some scenarios \cite{lundberg2017power,Sohanpal2025}. However, a key question remains: Once the C and L bands have been fully utilised, which bands should be added to maximise throughput whilst minimising the power consumption?
% Power consumption analysis is particularly important in fibre-scarce links, where additional throughput is required whilst maintaining energy efficiency. 
%Once the C and L bands have been exhausted, what is the next band that should be added to maximise transmission throughput whilst minimising the power use?

To answer this, in this work we measured the electrical-to-optical power conversion efficiency of various state-of-the-art DFAs covering the OESCL-bands. Building on our previous work \cite{Sohanpal2025} we have carried out measurements of O- and E-band BDFAs. We then used the integral GN model with launch power optimisation \cite{Jarmolovicius2024} to study the trade-off between throughput and energy-per-bit of various band combinations, showing that OESCL-band systems can achieve 2.98$\times$ throughput compared to CL-band systems for a +48\% increase in energy-per-bit.

 % for all band combinations across the OESCL-bands, highlighting potential avenues for the deployment of UWB in fibre-scarce systems.

\vspace{-0.15cm}
\section{OESCL amplification power efficiency measurements}
\vspace{-0.1cm}

\begin{figure}[b]

\begin{tikzpicture}[font=\footnotesize]

% \node[inner sep=0pt] (background) at (0,0) {\includegraphics[width=0.3\linewidth]{Figs/amplifierschematic.pdf}};

    \begin{scope}[shift={(0,-1.2)}]
    
    \begin{axis}
    [
    width=0.5\textwidth,height=5cm,
    grid=both,
    legend style={fill opacity=1, draw opacity=1, text opacity=1, at={(1.01,1)}, anchor=north west, draw=black, nodes={scale=0.7, transform shape}, legend columns=1},
    ylabel near ticks,
    xlabel near ticks,
    ylabel shift = -2 pt,
    xlabel shift = -2 pt,
    clip marker paths=true,
    ytick distance=1,
    % minor y tick num = 1,
    ymin=0,
    ymax=6.2,
    xmin=0,
    xmax=205,
    xlabel=Output power (mW),
    ylabel=Wallplug PCE (\%),
    clip mode=individual,
    mark size=1pt,
    label style={font=\footnotesize},
    tick label style={font=\footnotesize},
    legend cell align = left,
    ]
         
    \addplot[green1,thick,mark=*,mark options={fill=white}] table[x=xc,y=yc] {Data/pcedatac.tsv};
    \addlegendentry{C-EDFA}

    \addplot[orange1,thick,mark=square*,mark options={fill=white}] table[x=xl,y=yl] {Data/pcedatal.tsv};
    \addlegendentry{L-EDFA}
    
    \addplot[purple1,thick,mark=triangle*,mark size=1.5,mark options={fill=white}] table[x=xs,y=ys] {Data/pcedatas.tsv};
    \addlegendentry{S-TDFA}

    \addplot[lightgreen1,thick,dashed,mark=diamond*,mark size=1.5,mark options={solid}] table[x=xe_0,y=ye_0] {Data/pcedatae.tsv};
    \addlegendentry{E-BDFA (0 dBm input)}
    % \addplot[lightgreen1,thick,mark=triangle*,mark size=1.5] table[x=xe_2,y=ye_2] {Data/pcedatae.tsv};
    % \addlegendentry{BDFA (E-band, 2 dBm)}
    \addplot[lightgreen1,thick,mark=diamond*,mark size=1.5,mark options={fill=white}] table[x=xe_4,y=ye_4] {Data/pcedatae.tsv};
    \addlegendentry{E-BDFA (4 dBm input)}

    % \addplot[pink1,thick,dashed,mark=triangle*,mark size=1.5] table[x=xo_m4,y=yo_m4] {Data/pcedatao.tsv};
    % \addlegendentry{BDFA (O-band, -4 dBm)}
    
    \addplot[pink1,thick,dashed,mark=triangle*,mark size=1.5,mark options={solid}] table[x=xo_0,y=yo_0] {Data/pcedatao.tsv};
    \addlegendentry{O-BDFA (0 dBm input)}
    
    \addplot[pink1,thick,mark=triangle*,mark size=1.5,mark options={fill=white}] table[x=xo_4,y=yo_4] {Data/pcedatao.tsv};
    \addlegendentry{O-BDFA (4 dBm input)}

    \draw[gray] (52,0.25) rectangle (120,1.4);
    \draw[gray,densely dashed] (52,1.35) -- (58,2.1);
    \draw[gray,densely dashed] (120,1.35) -- (110,2.1);

    \draw[gray] (145,0.8) rectangle (203,1.5);
    \draw[gray,densely dashed] (145,1.5) -- (151,2.1);
    \draw[gray,densely dashed] (203,1.5) -- (180,2.1);

    % \node[at={(axis description cs:-1.15,0.97)},font=\normalsize,]{(a)};

    % \spy [black, draw, height = .8cm, width = 3cm, magnification = 2,
    % connect spies] on (2, -.9) in node [fill=white] at (3, -.2);
    % \spy[blue,size=2.5cm,magnification=1.8] on (axis description cs:0.05,0.9) in node at (axis description cs:0.5,0.5);
    
    \node[at={(axis description cs:0.05,0.9)},font=\normalsize,]{(a)};

    \node[at={(axis description cs:1.5,0.93)},font=\normalsize,]{(b)};
        
    \end{axis}
    \end{scope}

    \begin{scope}[shift={(1.8, -.02)}]
    \pgfplotsset{every tick label/.append style={font=\tiny}}
    \begin{axis}
    [
    width=3.2cm,height=2.3cm,
    minor y tick num = 2,
    ymin=0.3,
    ymax=1.35,
    xmin=52,
    xmax=120,
    ytick={0.4,0.8,1.2},
    xtick = {60,90,110},
    ylabel near ticks,
    xlabel near ticks,
    yticklabel style = {xshift=0.05cm},
    xticklabel style = {yshift=0.08cm},
    ]

    \draw[fill=white] (axis description cs:0,0) rectangle (axis description cs:1,1);

    \draw[gray,very thin] (50,1.2) -- (120,1.2);
    \draw[gray,very thin] (50,0.8) -- (120,0.8);
    \draw[gray,very thin] (50,0.4) -- (120,0.4);
    \draw[gray,very thin] (60,2) -- (60,0);
    \draw[gray,very thin] (90,2) -- (90,0);
    \draw[gray,very thin] (110,2) -- (110,0);
    
    % \draw[gray,very thin] (50,.8) -- (100,.8);
    % \draw[gray,very thin] (50,.6) -- (100,.6);
    % \draw[gray,very thin] (50,.4) -- (100,.4);
    % \draw[gray,very thin] (60,2) -- (60,0);
    % \draw[gray,very thin] (70,2) -- (70,0);
    % \draw[gray,very thin] (80,2) -- (80,0);
    \coordinate (subplot) at (axis description cs:0,0);

    \addplot[green1,thick,mark=*,mark options={fill=white}] table[x=xc,y=yc] {Data/pcedatac.tsv};

    \addplot[orange1,thick,mark=square*,mark options={fill=white}] table[x=xl,y=yl] {Data/pcedatal.tsv};
    
    \addplot[purple1,thick,mark=triangle*,mark size=1.5,mark options={fill=white}] table[x=xs,y=ys] {Data/pcedatas.tsv};

    \addplot[lightgreen1,thick,dashed,mark=triangle*,mark size=1.5,mark options={solid}] table[x=xe_0,y=ye_0] {Data/pcedatae.tsv};

    \addplot[lightgreen1,thick,mark=triangle*,mark size=1.5,mark options={fill=white}] table[x=xe_4,y=ye_4] {Data/pcedatae.tsv};

    \addplot[pink1,thick,dashed,mark=triangle*,mark size=1.5,mark options={solid}] table[x=xo_0,y=yo_0] {Data/pcedatao.tsv};
    
    \addplot[pink1,thick,mark=triangle*,mark size=1.5,mark options={fill=white}] table[x=xo_4,y=yo_4] {Data/pcedatao.tsv};

    \end{axis}
    \end{scope}

    \begin{scope}[shift={(4.7, -0.02)}]
    \pgfplotsset{every tick label/.append style={font=\tiny}}
    \begin{axis}
    [
    width=2.5cm,height=2.3cm,
    minor y tick num = 2,
    ymin=1.15,
    ymax=1.35,
    xmin=150,
    xmax=210,
    ytick={1.2,1.3},
    xtick = {160,200},
    ylabel near ticks,
    xlabel near ticks,
    yticklabel style = {xshift=0.05cm},
    xticklabel style = {yshift=0.08cm},
    ]

    \draw[fill=white] (axis description cs:0,0) rectangle (axis description cs:1,1);

    \draw[gray,very thin] (150,1.2) -- (220,1.2);
    \draw[gray,very thin] (150,1.3) -- (220,1.3);
    \draw[gray,very thin] (160,2) -- (160,0);
    \draw[gray,very thin] (200,2) -- (200,0);
    
    % \draw[gray,very thin] (50,.8) -- (100,.8);
    % \draw[gray,very thin] (50,.6) -- (100,.6);
    % \draw[gray,very thin] (50,.4) -- (100,.4);
    % \draw[gray,very thin] (60,2) -- (60,0);
    % \draw[gray,very thin] (70,2) -- (70,0);
    % \draw[gray,very thin] (80,2) -- (80,0);
    \coordinate (subplot) at (axis description cs:0,0);

    \addplot[lightgreen1,thick,dashed,mark=triangle*,mark size=1.5,mark options={solid}] table[x=xe_0,y=ye_0] {Data/pcedatae.tsv};

    \addplot[lightgreen1,thick,mark=triangle*,mark size=1.5,mark options={fill=white}] table[x=xe_4,y=ye_4] {Data/pcedatae.tsv};

    \end{axis}
    \end{scope}

    \begin{scope}[shift={(10.3,-1.2)}]
        \newcommand\fibreFigWidth{.4\textwidth}
        \newcommand\fibreFigHeight{5cm}
    
        \input{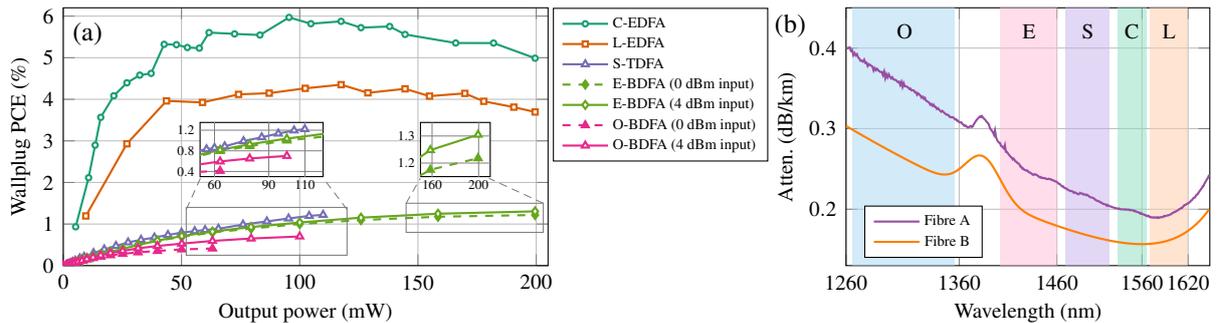}
    \end{scope}

\end{tikzpicture}

\vspace{-1.5em}
\caption{(a) Measured wallplug PCE for various doped-fibre amplifiers. (b) Attenuation profiles used in the integral GN model.}
\label{fig:fig1}

\end{figure}

The DFA power consumption for any given wavelength band can be defined as $P_{\text{elec}} = (P_{\text{out}} - P_{\text{in}})/\eta $, where $P_{\text{out}}$ and $P_{\text{in}}$ are the total output and input optical powers respectively, and $\eta$ is the wallplug power conversion efficiency (PCE). The PCE accounts for the power consumption of the drivers, the electrical-to-optical conversion efficiency of the pump laser diodes, and the optical-to-optical conversion efficiency of the doped fibre. For simplicity, we ignore any additional power consumption overheads (e.g. fans, control and monitoring etc). 

To estimate the power consumption of various DFAs, we first determined $\eta$ for each DFA, using the methodology in \cite{Sohanpal2025}. For each band, we generated spectrally-shaped ASE (SS-ASE) to emulate WDM signals for a given total input power and launched it into the DFA under test. The SS-ASE was shaped and flattened to cover the entire DFA gain bandwidth in all bands except the E-band, where a spectral processor was not available and unshaped ASE was used. The pump currents were swept whilst the output power and wallplug electrical power consumption was measured using a commercial power monitor. To remove power overheads, we subtracted any ambient power draw when there was no pump current. Five state-of-the-art DFAs were measured:  C-band  (1530nm-1565nm) and L-band (1570nm-1620nm) EDFAs and an E-band BDFA (1400nm-1460nm) all from Amonics, an S-band TDFA (1470nm-1520nm) from Fiberlabs and a custom non-commercial O-band BDFA (1265nm-1355nm) from Lightera \cite{Mikhailov2022}. The noise figures of the O-, E-, S-, C- and L-band amplifiers were 5~dB, 6.5~dB, 7~dB, 5~dB and 6~dB, respectively. All DFAs have been used previously in various UWB transmission experiments \cite{Puttnam2025,Aparecido2025,Yang2025ecoc,Yang2025,Luis2025}. 

The wallplug PCE results are shown in Fig.~\ref{fig:fig1}(a). For both EDFAs and the TDFA, it was observed that the PCE was invariant with input power, and 2~dBm total input power was used to achieve PCEs of approximately 5\% (C-EDFA), 3.7\% (L-EDFA) and 1.2\% (TDFA). However, the O- and E-band BDFA PCEs vary as a function of input power, with the E-BDFA PCE varying from 1.2\% to 1.3\% and the O-BDFA PCE varying from 0.4\% to 0.7\% for 0~dBm and 4~dBm input power respectively. Note that these PCEs are for commercial and research DFAs and are not optimised for a given link, thus they reflect the worst-case scenario for future deployed systems.

To evaluate the power consumption of OESCL-band systems, we used the integral Gaussian Noise model in the presence of inter-channel stimulated Raman scattering (ISRS) to calculate the received power profile and SNR for various UWB scenarios \cite{Jarmolovicius2024}. Here, we considered only one dedicated amplifier per band. For each band, we modelled 140~GBd dual-polarisation Gaussian-constellation channels with a 150~GHz spacing and transceiver back-to-back SNR of 20~dB. We restricted the transmitted wavelengths to match the measured amplifiers, resulting in 277~channels between 1265~nm and 1620~nm. In addition, we implemented segmented launch power optimisation for each UWB scenario to maximise the Shannon capacity by mitigating ISRS and fibre nonlinearity, as described in \cite{Jarmolovicius2024}. Two single-mode fibre attenuation profiles were considered: a measured G.652.D-compliant fibre profile from a deployed system (Fibre A) \cite{Yang2025} and a simulated low-loss fibre (Fibre B), with losses of 0.20~dB/km and 0.15~dB/km at 1550~nm respectively, shown in Fig.~\ref{fig:fig1}(b). Both fibres were modelled with chromatic  dispersion values of -2.4~ps/(nm$\cdot$km), 4.5~ps/(nm$\cdot$km) and 16.9~ps/(nm$\cdot$km) at 1260~nm, 1360~nm and 1550~nm, respectively and nonlinear coefficient of 1.97~W$^{-1}$km$^{-1}$ and 1.27~W$^{-1}$km$^{-1}$ at 1310~nm and 1550~nm respectively. The span length of each fibre link was 80~km.

\vspace{-0.15cm}
\section{Results and discussion}
\vspace{-0.15cm}

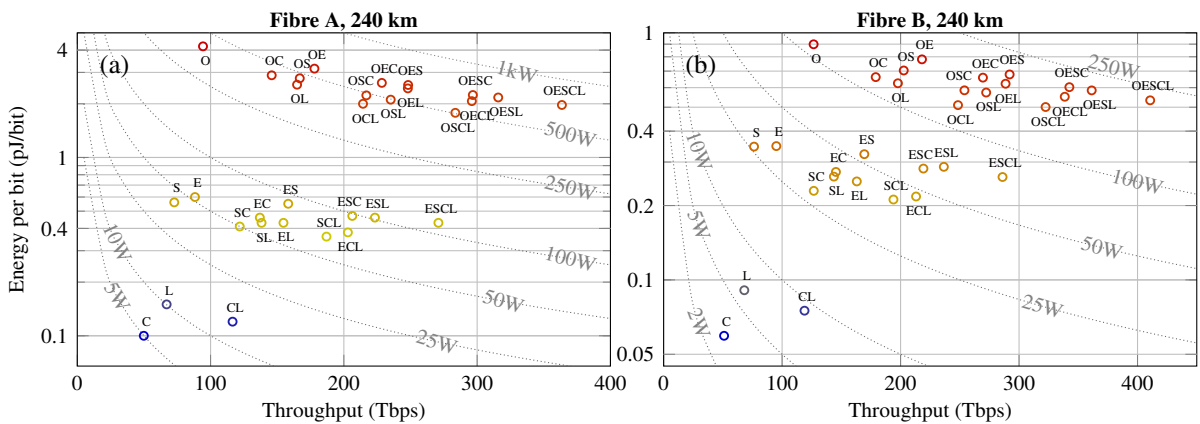
\begin{figure}[b]
\begin{tikzpicture}[font=\footnotesize]

    \begin{groupplot}
    [
    width=0.54\textwidth,height=6cm,
    grid=both,
    ylabel near ticks,
    xlabel near ticks,
    ylabel shift = -2 pt,
    xlabel shift = -2 pt,
    clip marker paths=true,
    % ytick distance=1,
    % % minor y tick num = 1,
    % ymin=0,
    % ymax=6.2,
    % xmin=0,
    % xmax=205,
    ymode=log,
    xlabel=Throughput (Tbps),
    ylabel=Energy per bit (pJ/bit),
    % clip mode=individual,
    % mark size=1pt,
    label style={font=\footnotesize},
    tick label style={font=\footnotesize},
    grid=both,
    group style={group size=2 by 1,xlabels at=edge bottom,ylabels at=edge left,horizontal sep=0.7cm},
    title style={at={(0.5,0.93)},font=\bfseries\footnotesize},
    ]

    \nextgroupplot[
    title = {Fibre A, 240~km},
    ymax = 5,
    ytick={0.1,1,10},
    yticklabels={0.1,1,10},
    extra y ticks = {0.4,4},
    extra y tick labels={0.4,4},
    xmin=0,
    xmax=400,
    % yticklabels={0.1,1,4},
    % ytick={0.1,0.2,0.5,1,2,4},
    % yticklabels={0.1,0.2,0.5,1,2,4},
    % colormap={bw}{
    %     red(0cm)=(0);
    %     blue(1cm)=(1);
    % }
    ]

    \draw[scale=1, domain=5:400, smooth, variable=\x, gray, densely dotted] plot ({\x}, {1/\x*1000});    
    \draw[scale=1, domain=5:400, smooth, variable=\x, gray, densely dotted] plot ({\x}, {1/\x*500});
    \draw[scale=1, domain=5:400, smooth, variable=\x, gray1, densely dotted] plot ({\x}, {1/\x*250});
    \draw[scale=1, domain=5:400, smooth, variable=\x, gray1, densely dotted] plot ({\x}, {1/\x*100});
    \draw[scale=1, domain=5:400, smooth, variable=\x, gray1, densely dotted] plot ({\x}, {1/\x*50});
    \draw[scale=1, domain=5:400, smooth, variable=\x, gray1, densely dotted] plot ({\x}, {1/\x*25});
    \draw[scale=1, domain=5:400, smooth, variable=\x, gray1, densely dotted] plot ({\x}, {1/\x*10});
    \draw[scale=1, domain=5:400, smooth, variable=\x, gray1, densely dotted] plot ({\x}, {1/\x*5});
    \node[rotate=0,gray2,rotate=-60] at (30,1/30*5) {5W};
    \node[rotate=0,gray2,rotate=-65] at (30,1/30*10) {10W};
    \node[rotate=0,gray2,rotate=-17] at (270,1/270*25) {25W};
    \node[rotate=0,gray2,rotate=-17] at (320,1/320*50) {50W};
    \node[rotate=0,gray2,rotate=-15] at (370,1/370*100) {100W};
    \node[rotate=0,gray2,rotate=-15] at (370,1/370*250) {250W};
    \node[rotate=0,gray2,rotate=-15] at (370,1/370*500) {500W};
    \node[rotate=0,gray2,rotate=-15] at (330,1/330*1000) {1kW};

    \node[at={(axis description cs:0.07,0.9)},font=\normalsize,]{(a)};

    \addplot[only marks, 
    % scatter src=1/x*y,
    mark=o, 
    mark size = 1.5, 
    line width=0.7pt,
    % node near coord style={color=black,anchor=south west, font=\tiny},
    % visualization depends on={\thisrow{label} \as \annotvalue},
    % nodes near coords,
    % point meta=explicit symbolic,
     % Node labels
     % scatter src = explicit symbolic,
    % visualization depends on = {\thisrow{pjperbit_amponly} \as \perpointmarksize},
     % scatter/@pre marker code/.append style={/tikz/mark size = \perpointmarksize},
    visualization depends on = {value \thisrow{label} \as \perpointlabel},
    visualization depends on = {value \thisrow{anchor} \as \perpointanchor},
    visualization depends on = {value \thisrow{xshift} \as \perpointxshift},
     nodes near coords* = {\perpointlabel},
     every node near coord/.append style={
        font = \tiny,
        color=black,
        anchor=\perpointanchor,
        yshift=0,
        xshift=\perpointxshift,
     }
    ] table[x=capacity,y=pjperbit_amponly,meta=label] {Data/measfibre3.tsv}; %\capdata;

    % \addplot[domain=0.1:10,
    %         domain y=0:100,
    %         contour gnuplot={levels={-1,1},labels=false},
    %         thick,samples=50,samples y=50,
    %     ] {x^2-x*y};

    % \addplot gnuplot[raw gnuplot,thick,mark=none,blue,very thin] {
    %     unset surface;
    %     set cntrparam levels discrete 0,50;
    %     set contour;
    %     set yrange [0.05:10];
    %     splot [x=0:400] y*x*10;
    % };

        \nextgroupplot[
    title = {Fibre B, 240~km},
    ymax = 1,
    ytick={0.1,1},
    yticklabels={0.1,1},
    extra y ticks = {0.05,0.2,0.4},
    extra y tick labels={0.05,0.2,0.4},
    xmin=0,
    xmax=450,
    ]

    \node[at={(axis description cs:0.07,0.9)},font=\normalsize,]{(b)};

    \draw[scale=1, domain=5:450, smooth, variable=\x, gray1, densely dotted] plot ({\x}, {1/\x*250});
    \draw[scale=1, domain=5:450, smooth, variable=\x, gray1, densely dotted] plot ({\x}, {1/\x*100});
    \draw[scale=1, domain=5:450, smooth, variable=\x, gray1, densely dotted] plot ({\x}, {1/\x*50});
    \draw[scale=1, domain=5:450, smooth, variable=\x, gray1, densely dotted] plot ({\x}, {1/\x*25});
    \draw[scale=1, domain=5:450, smooth, variable=\x, gray1, densely dotted] plot ({\x}, {1/\x*10});
    \draw[scale=1, domain=5:450, smooth, variable=\x, gray1, densely dotted] plot ({\x}, {1/\x*5});
    \draw[scale=1, domain=5:450, smooth, variable=\x, gray1, densely dotted] plot ({\x}, {1/\x*2});
    \node[rotate=0,gray2,rotate=-70] at (30,2/30*1) {2W};
    \node[rotate=0,gray2,rotate=-70] at (30,1/30*5) {5W};
    \node[rotate=0,gray2,rotate=-70] at (30,1/30*10) {10W};
    \node[rotate=0,gray2,rotate=-17] at (320,1/320*25) {25W};
    \node[rotate=0,gray2,rotate=-15] at (370,1/370*50) {50W};
    \node[rotate=0,gray2,rotate=-15] at (400,1/400*100) {100W};
    \node[rotate=0,gray2,rotate=-15] at (380,1/380*250+.07) {250W};

    \addplot[only marks, 
    % scatter src=1/x*y,
    mark=o, 
    mark size = 1.5, 
    line width=0.7pt,
    visualization depends on = {value \thisrow{label} \as \perpointlabel},
    visualization depends on = {value \thisrow{anchor} \as \perpointanchor},
    visualization depends on = {value \thisrow{xshift} \as \perpointxshift},
     nodes near coords* = {\perpointlabel},
     every node near coord/.append style={
        font = \tiny,
        color=black,
        anchor=\perpointanchor,
        yshift=0,
        xshift=\perpointxshift,
     }
    ] table[x=capacity,y=pjperbit_amponly,meta=label] {Data/simfibre3.tsv}; %\capdata;

    \end{groupplot}

\end{tikzpicture}

\vspace{-1em}
\caption{Amplification energy per bit versus throughput for all 31 optimised OESCL-band combinations using (a) Fibre A and (b) Fibre B for 3 spans (240~km). Dotted lines show regions of fixed amplifier power consumption.}
\label{fig:fig2}
\end{figure}

\begin{figure}[t]
\begin{tikzpicture}[font=\footnotesize]

    \begin{groupplot}
    [
    width=0.54\textwidth,height=6cm,
    grid=both,
    ylabel near ticks,
    xlabel near ticks,
    ylabel shift = -2 pt,
    xlabel shift = -2 pt,
    clip marker paths=true,
    % ytick distance=1,
    % % minor y tick num = 1,
    % ymin=0,
    % ymax=6.2,
    % xmin=0,
    % xmax=205,
    ymode=log,
    xlabel=Throughput (Tbps),
    ylabel=Energy per bit (pJ/bit),
    % clip mode=individual,
    % mark size=1pt,
    label style={font=\footnotesize},
    tick label style={font=\footnotesize},
    grid=both,
    group style={group size=2 by 1,xlabels at=edge bottom,ylabels at=edge left,horizontal sep=0.7cm},
    title style={at={(0.5,0.93)},font=\bfseries\footnotesize},
    ]

    \nextgroupplot[
    title = {Fibre B, 1040~km, amplifier only},
    ymax = 6,
    ytick={0.1,1,10},
    yticklabels={0.1,1,10},
    extra y ticks = {0.4,2,4},
    extra y tick labels={0.4,2,4},
    xmin=0,
    xmax=320,
    ]

    \draw[scale=1, domain=5:400, smooth, variable=\x, gray, densely dotted] plot ({\x}, {1/\x*1000});    
    \draw[scale=1, domain=5:400, smooth, variable=\x, gray, densely dotted] plot ({\x}, {1/\x*500});
    \draw[scale=1, domain=5:400, smooth, variable=\x, gray1, densely dotted] plot ({\x}, {1/\x*250});
    \draw[scale=1, domain=5:400, smooth, variable=\x, gray1, densely dotted] plot ({\x}, {1/\x*100});
    \draw[scale=1, domain=5:400, smooth, variable=\x, gray1, densely dotted] plot ({\x}, {1/\x*50});
    \draw[scale=1, domain=5:400, smooth, variable=\x, gray1, densely dotted] plot ({\x}, {1/\x*25});
    \draw[scale=1, domain=5:400, smooth, variable=\x, gray1, densely dotted] plot ({\x}, {1/\x*10});
    \node[rotate=0,gray2,rotate=-65] at (30,1/30*10) {10W};
    \node[rotate=0,gray2,rotate=-65] at (30,1/30*25) {25W};
    \node[rotate=0,gray2,rotate=-17] at (170,1/170*50) {50W};
    \node[rotate=0,gray2,rotate=-15] at (220,1/220*100) {100W};
    \node[rotate=0,gray2,rotate=-15] at (260,1/260*250) {250W};
    \node[rotate=0,gray2,rotate=-15] at (280,1/280*500) {500W};
    \node[rotate=0,gray2,rotate=-15] at (275,1/275*1000) {1kW};

    \node[at={(axis description cs:0.07,0.9)},font=\normalsize,]{(a)};

    \addplot[only marks, 
    % scatter src=1/x*y,
    mark=o, 
    mark size = 1.5, 
    line width=0.7pt,
    % node near coord style={color=black,anchor=south west, font=\tiny},
    % visualization depends on={\thisrow{label} \as \annotvalue},
    % nodes near coords,
    % point meta=explicit symbolic,
     % Node labels
     % scatter src = explicit symbolic,
    % visualization depends on = {\thisrow{pjperbit_amponly} \as \perpointmarksize},
     % scatter/@pre marker code/.append style={/tikz/mark size = \perpointmarksize},
    visualization depends on = {value \thisrow{label} \as \perpointlabel},
    visualization depends on = {value \thisrow{anchor} \as \perpointanchor},
    visualization depends on = {value \thisrow{xshift} \as \perpointxshift},
     nodes near coords* = {\perpointlabel},
     every node near coord/.append style={
        font = \tiny,
        color=black,
        anchor=\perpointanchor,
        yshift=0,
        xshift=\perpointxshift,
     }
    ] table[x=capacity,y=pjperbit_amponly,meta=label] {Data/simfibre13.tsv}; %\capdata;

    % \addplot[domain=0.1:10,
    %         domain y=0:100,
    %         contour gnuplot={levels={-1,1},labels=false},
    %         thick,samples=50,samples y=50,
    %     ] {x^2-x*y};

    % \addplot gnuplot[raw gnuplot,thick,mark=none,blue,very thin] {
    %     unset surface;
    %     set cntrparam levels discrete 0,50;
    %     set contour;
    %     set yrange [0.05:10];
    %     splot [x=0:400] y*x*10;
    % };

        \nextgroupplot[
    title = {Fibre B, 1040~km, amplifier + transceiver},
    ymax = 40,
    ytick={20,30},
    yticklabels={20,30},
    extra y ticks = {15,25,35,40},
    extra y tick labels={15,25,35,40},
    xmin=0,
    xmax=320,
    ]

    \node[at={(axis description cs:0.07,0.9)},font=\normalsize,]{(b)};

    \draw[scale=1, domain=5:450, smooth, variable=\x, gray1, densely dotted] plot ({\x}, {1/\x*500});
    \draw[scale=1, domain=5:450, smooth, variable=\x, gray1, densely dotted] plot ({\x}, {1/\x*1000});
    \draw[scale=1, domain=5:450, smooth, variable=\x, gray1, densely dotted] plot ({\x}, {1/\x*2500});
    \draw[scale=1, domain=5:450, smooth, variable=\x, gray1, densely dotted] plot ({\x}, {1/\x*5000});
    \draw[scale=1, domain=5:450, smooth, variable=\x, gray1, densely dotted] plot ({\x}, {1/\x*10000});
    % \draw[scale=1, domain=5:450, smooth, variable=\x, gray1, densely dotted] plot ({\x}, {1/\x*5});
    % \draw[scale=1, domain=5:450, smooth, variable=\x, gray1, densely dotted] plot ({\x}, {1/\x*2});
    \node[rotate=0,gray2,rotate=-85] at (22,1/22*500) {500W};
    \node[rotate=0,gray2,rotate=-80] at (40,1/40*1000) {1kW};
    \node[rotate=0,gray2,rotate=-70] at (90,1/90*2500) {2.5kW};
    \node[rotate=0,gray2,rotate=-35] at (280,1/280*5000) {5kW};
    \node[rotate=0,gray2,rotate=-30] at (280,1/280*10000) {10kW};
    % \node[rotate=0,gray2,rotate=-15] at (370,1/370*50) {50W};
    % \node[rotate=0,gray2,rotate=-15] at (400,1/400*100) {100W};
    % \node[rotate=0,gray2,rotate=-15] at (380,1/380*250+.07) {250W};

    \draw [stealth-stealth,line width = 1] (axis cs:106,16.15)--(axis cs:307,16.15) node[midway,fill=white,inner sep=1pt,font=\footnotesize,anchor=east] {2.98$\times$};

    \draw [stealth-stealth,line width = 1] (axis cs:308,16.15)--(axis cs:308,23.6) node[midway,inner sep=1pt,font=\footnotesize,anchor=east] {+48\%};

    \addplot[only marks, 
    % scatter src=-1/x*y,
    mark=o, 
    mark size = 1.5, 
    line width=0.7pt,
    visualization depends on = {value \thisrow{label} \as \perpointlabel},
    visualization depends on = {value \thisrow{anchor} \as \perpointanchor},
    visualization depends on = {value \thisrow{xshift} \as \perpointxshift},
     nodes near coords* = {\perpointlabel},
     every node near coord/.append style={
        font = \tiny,
        color=black,
        anchor=\perpointanchor,
        yshift=0,
        xshift=\perpointxshift,
     }
    ] table[x=capacity,y=pjperbit_withtrx,meta=label] {Data/simfibre13.tsv}; %\capdata;
    \end{groupplot}

\end{tikzpicture}

\vspace{-1em}
\caption{Energy per bit versus throughput using Fibre B for 13 spans (1040~km) for (a) amplifier power consumption only and  (b) both amplifier and transceiver. Dotted lines show regions of fixed amplifier power consumption.}
\label{fig:fig3}
\vspace{-.3cm}
\end{figure}
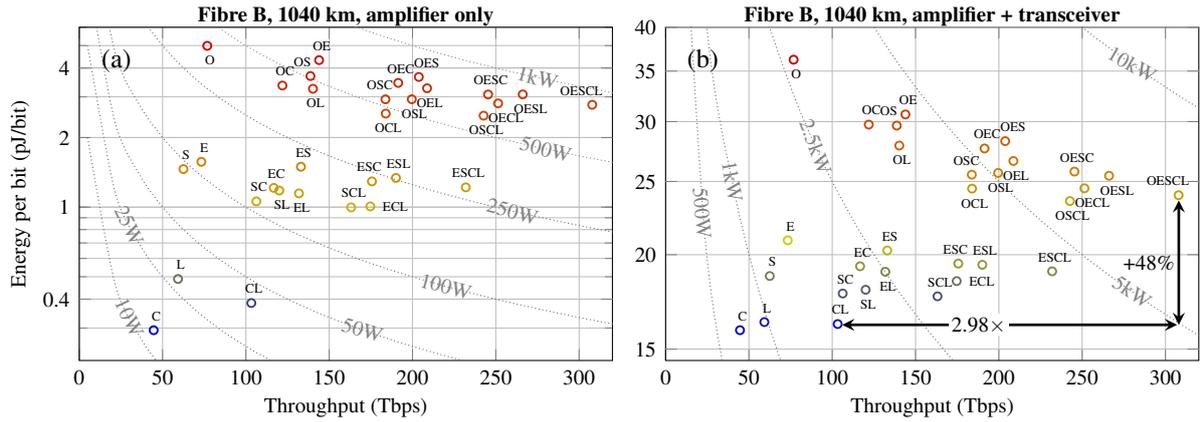

Using the integral GN model and the measured amplifier PCEs, we modelled all unique band combinations within the set \{O,E,S,C,L\} whilst optimising the launch power of each scenario. This was repeated for both fibre variants and for transmission distances of 3 and 13 spans (240~km and 1040~km respectively). There are three main factors that determine the energy-per-bit for each band: the amplifier PCE, the bandwidth and the fibre loss. Wider bands will require higher optical output power than narrower bands at equivalent launch power per channel. Further, bands at higher attenuation wavelengths will also require higher launch power to maintain sufficient received OSNR. The O- and E-bands will typically require higher electrical power than the C- and L-bands, even assuming identical amplifier PCEs. This is further exacerbated by the BDFA PCE increasing with input power -- higher fibre attenuation at short wavelengths will lead to lower received power at the end of each span, leading to reduced PCE and even higher power consumption. The impact of fibre nonlinearity, including ISRS, on signal distortion and channel power profiles will also affect the energy-per-bit. 

Fig.~\ref{fig:fig2} shows the amplification energy-per-bit versus throughput for all 31 band combinations, ranging from individual bands to the entire OESCL-band spectrum, for the standard-loss Fibre A and the low-loss Fibre B for 240~km. For both fibres the EDFA is the most power efficient, leading to the lowest energy-per-bit of less than 0.2~pJ/bit for the C-, L- and CL-bands. In this case, the throughput after 240~km is mainly limited by the CL bandwidth and the transceiver SNR. Comparing both fibre profiles, the lower attenuation of Fibre B leads to improved energy-per-bit for all band combinations. This is most significant for the O- and E-bands, which both benefit from throughput gains as well as considerable reductions in energy-per-bit. The O-band-only throughput increases by over 34\% from 94~Tbps to 127~Tbps whilst the energy-per-bit reduces from 4.2~pJ/bit to 0.9~pJ/bit, highlighting the importance of low-loss fibre for energy efficient operation in BDFA-based O- and E-band systems.

Fig.~\ref{fig:fig3} shows the energy-per-bit versus throughput for Fibre B at 13 spans (1040~km) with and without a transceiver power contribution. Here, we estimated a pluggable transceiver power consumption of 24~W per 140~GBd channel, assuming the worst case scenario of maximum power draw across all transceivers. It is clear from Fig.~\ref{fig:fig3}(b) that the transceiver power consumption dominates over that of the amplifiers alone, leading to the energy-per-bit becoming proportional to the number of channels (i.e. the bandwidth) rather than the amplifier characteristics. Considering Fig.~\ref{fig:fig3}(a), %the optimal bands (largest throughput for smallest energy-per-bit) are \{C, CL, ECL, ESCL, OESCL\}. 
the optimal order of band deployment (highest throughput for lowest energy-per-bit) is C$\to$CL$\to$ECL$\to$ESCL$\to$OESCL. The E- and S-band combinations offer similar energy-per-bit, but the E-band adds higher throughput. The S-band throughput would benefit considerably from DRA, thus this finding may change when DRA (or an alternative amplification technology, e.g. S-band BDFA \cite{Donodin2025Sband}) is employed. Considering Fig.~\ref{fig:fig3}(b), taking transceiver power consumption into account, the optimal order of band deployment is %\{C, CL, SCL, ESCL, OESCL\}
C$\to$CL$\to$SCL$\to$ESCL$\to$OESCL, where SCL slightly improves upon ECL energy-per-bit as it employs fewer channels. In multi-band systems, the energy-per-bit of all bands is averaged -- more efficient bands (e.g. C-band) ameliorate the less efficient bands (e.g. O-band) when transmitted together. In single- or dual-band systems, it is always most energy efficient to use EDFA-amplified bands only.

\vspace{-0.15cm}
\section{Conclusion}
\vspace{-0.15cm}

We investigated the energy efficiency of various state-of-the-art OESCL-band amplifiers. Using the integral GN model, we showed that OESCL systems can achieve almost a factor of 3 higher throughput than CL systems for only a 48\% energy-per-bit increase at 1000~km.

% \vspace{0.2cm}
\noindent\textbf{Acknowledgements:} This work was supported by EPSRC grants EP/R035342/1 TRANSNET% (Transforming networks - building an intelligent optical infrastructure)
, EP/W015714/1 EWOC% (Extremely Wideband Optical Fibre Communication Systems)
, and EP/V007734/1. The authors thank Dr Aleksandr Donodin (AiPT) for valuable discussions.

% \vspace{-0.2cm}
% \bibliographystyle{opticajnl}
% \bibliography{mybib}

\vspace{-0.2cm}
\section*{References}
\vspace{-0.6cm}
\begin{multicols}{2}
\bibliographystyle{opticajnl}
\bibliography{mybib}
\end{multicols}

\end{document}